\documentclass[global,twocolumn,final]{svjour}
\usepackage{graphics}
\usepackage{amssymb,amsmath,latexsym,amscd,amsfonts,latexsym,mathtools,mathrsfs,array,bm}

\usepackage[usenames,dvipsnames]{xcolor}
\usepackage{graphicx}
\usepackage[caption=false,font=footnotesize]{subfig}
\usepackage{lineno,hyperref}

\usepackage{soul}

\newcommand{\ud}{\,\mathrm{d}}
\begin{document}
\title{Optical wave evolution due to interaction with elastic wave in a phoxonic crystal slab waveguide}

\author{Mohammad Hasan Aram\inst{1} \and Sina Khorasani\inst{1}% etc
\thanks{\emph{Present address:} \'Ecole Polytechnique F\'ed\'erale de Lausanne,
CH-1015 Lausanne, Switzerland}%
}                     % Do not remove

\institute{School of Electrical Engineering, Sharif University of Technology, Azadi Ave., Tehran, Iran. , \email{mharam@ee.sharif.edu}}
\date{Received: date / Revised version: date}
\maketitle
\begin{abstract}
Phoxonic crystal as a means of guiding and confining electromagnetic and elastic waves has already attracted attentions. Lack of exact knowledge on how these two types of waves interact inside this crystal and how electromagnetic wave evolves through this interaction has increased this field complexity. Here we explain how an elastic wave affects an electromagnetic wave through photo-elasticity and interface displacement mechanisms in a phoxonic crystal slab waveguide. We obtain a master equation which can describe electromagnetic wave evolution. In this equation we define a coupling parameter and calculate its value for different modes of electromagnetic and elastic waves and show it vanishes for some types of modes . Finally we solve the master equation for a typical phoxonic crystal slab waveguide and illustrate the electromagnetic wave evolution. 
\end{abstract}
\section{Introduction}\label{intro}
About a century ago, French physicist, L\`{e}on Brillouin predicted diffraction of electromagnetic wave by elastic wave \cite{Quate}. Since then this phenomenon and generally interaction of electromagnetic and acoustic waves has attracted much attention and found many applications. At first these applications were limited to measuring elastic constants \cite{Vacher,Ferrari,Cusack} and finding thermal phonon distributions \cite{Toshio,Bao} of different materials. This phenomenon then got used for determining sound velocity \cite{Zha} and attenuation \cite{Pine} in various materials. It was also used to create coherent sound by means of intense laser beams \cite{Chiao}. In recent years, new ideas have been proposed to apply interaction of electromagnetic and elastic waves. These include classical and quantum mechanical aspects of this interaction. For instance, combined electromagnetic and elastic waves are used for detecting non-metallic buried objects \cite{Scott}, or optical isolation can be created by means of elastic wave in an optical waveguide \cite{Yu}. This means elastic waves cause non-reciprocity which is crucial in integrated photonic circuits \cite{Shen,Ruesink,Hafezi}. Enhancement of acousto-optic interaction in a phoxonic cavity \cite{Rong,Li} makes laser cooling feasible which can empty the cavity of any phonon \cite{Jasper,Aspelmeyer,Teufel,Rae}. This enables us to observe some quantum phenomena which little amount of thermal noise can disturb them. Active photonic crystal is another achievement of employing elastic waves \cite{Fuhrmann,Lima}. These crystals can be tuned for different purposes. Very precise displacement sensing may be its most interesting and effective new application. Using that interaction one can measure position beyond the standard quantum limit \cite{Schliesser,Anetsberger}.

\begin{figure}[!ht]
\centering
{\includegraphics[width=3.4in]{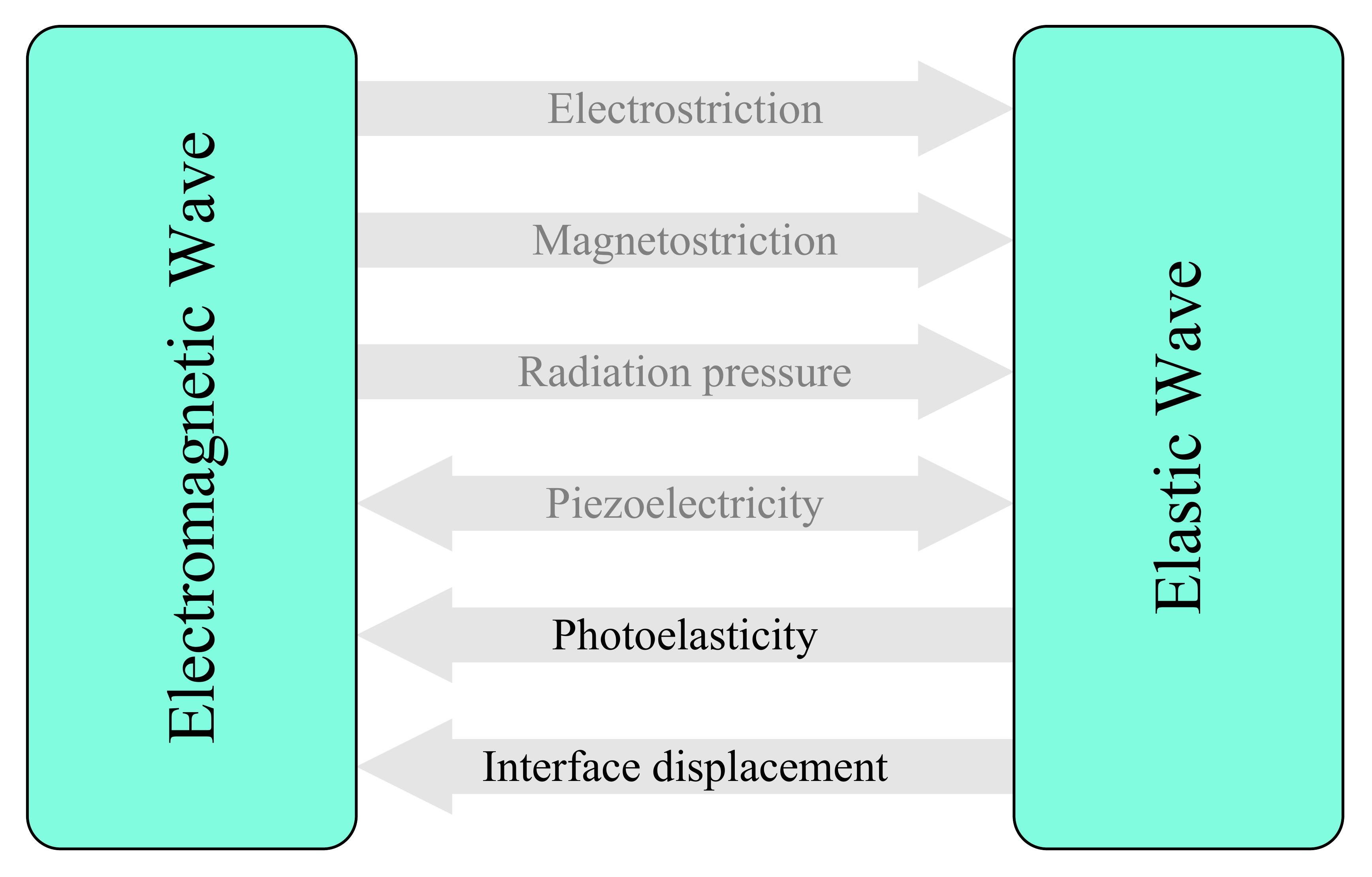}}
\caption{The schematic diagram which illustrates coupling mechanisms between elastic and electromagnetic waves. In this paper the last 2 are supposed to be considerable.}
\label{F0}
\end{figure}

Different mechanisms may be responsible for interaction of electromagnetic and elastic waves, among them are electrostriction, magnetostriction, radiation pressure, piezoelectricity, photoelasticity, and interface displacement \cite{Pennec3}. The first three are mechanisms through which optical waves affect elastic waves. These are not the subject of this paper because we assume the electromagnetic wave power to be very low. We also ignore piezoelectricity since no piezoelectric material is supposed to be present in our problem. If the elastic wave is strong enough, the last two mechanisms which are responsible for the effect of elastic wave on optical wave will become considerable. Photoelasticity causes the permittivity to change due to changes in material strain. In environments consisting of two or more materials with distinct boundaries, elastic wave can displace these boundaries and cause electromagnetic wave to encounter alternating environment. These two mechanisms can reinforce or weaken each other \cite{Rolland,Dupont}, that is their effects on electromagnetic wave are not necessarily cooperative. Fig. \ref{F0} illustrates these coupling mechanisms in a schematic. 

Phoxonic crystal is a good environment to experience both mechanisms. These crystals which can simultaneously have both photonic and phononic band gaps in their band structures, may be utilized to construct an optomechanical cavity or waveguide \cite{Mohammadi2010,Laude2011}. Since fields are confined inside these structures, we expect an enhanced interaction between optical and mechanical waves inside them \cite{Rong,ElJallal2013}.

To the best of our knowledge, there is no precise description of this interaction. In other words, how electromagnetic wave evolves in time during interaction with an elastic wave via photoelasticity and interface displacement mechanisms has not been studied well yet. In this paper we first extract a master equation which can describe this evolution inside a phoxonic waveguide. Then we analyze the coupling strength between different optical and mechanical modes. Finally optical wave dynamics is illustrated by solving the master equation.

\section{Problem statement} \label{S2}
\begin{figure*}[!ht]
\centering
{\includegraphics[width=5.02in]{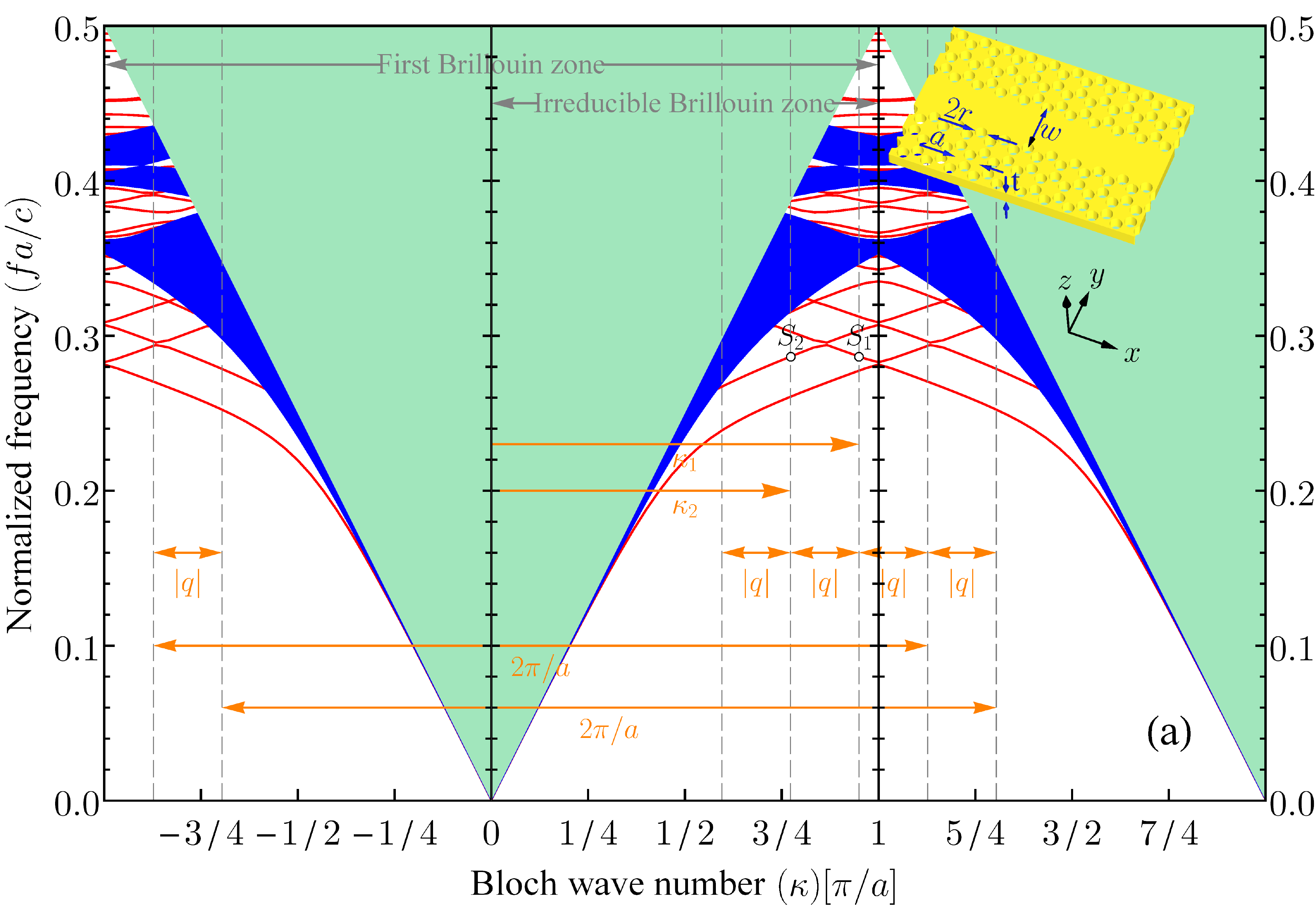}\label{F1a}}
\hspace{0pt}
{\includegraphics[width=1.895in]{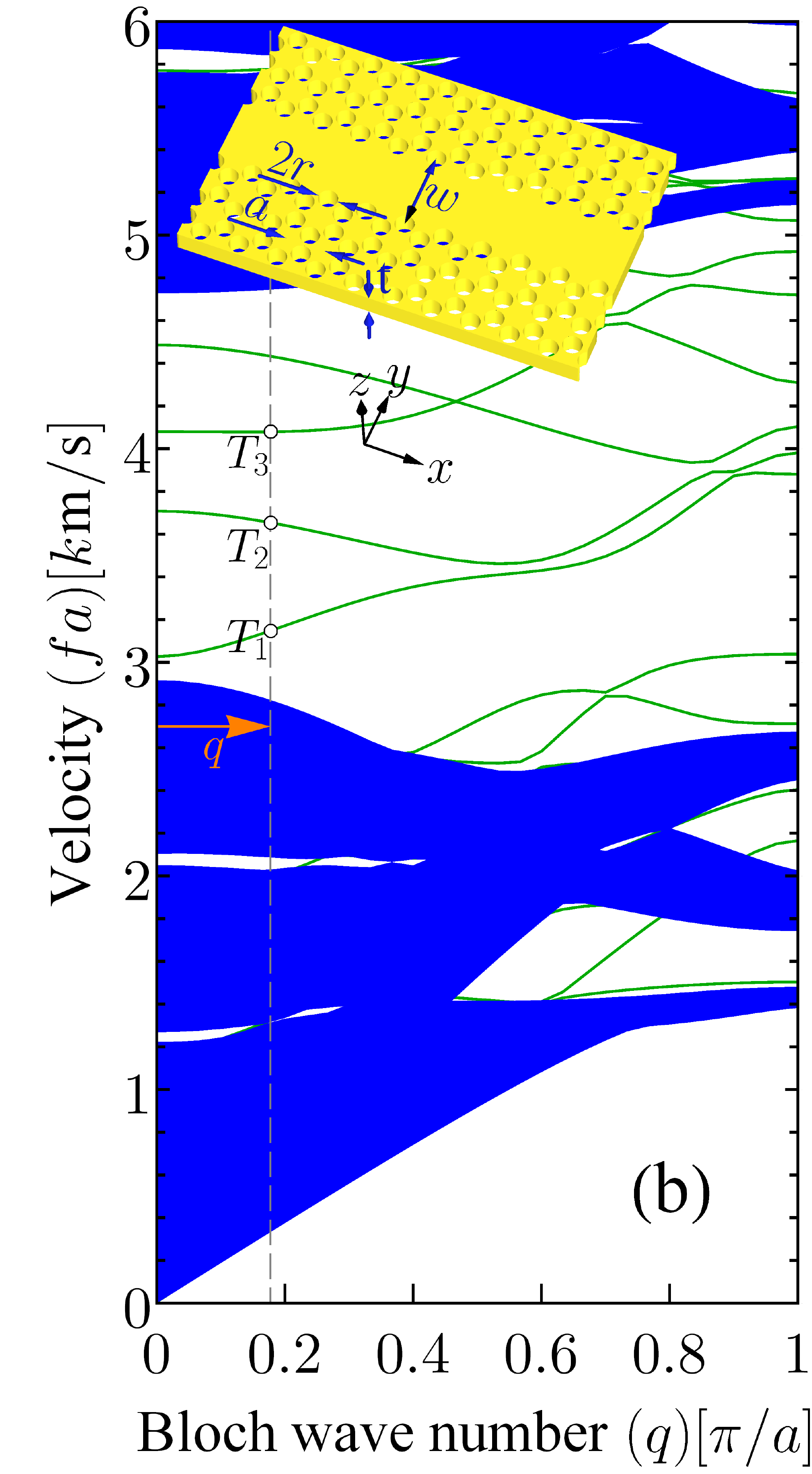}\label{F1b}} 
\caption{Band structure of a phoxonic crystal slab waveguide for (a) photonic TM-like and (b) phononic HD-like modes. The waveguide shown in the insets of figures is created by introducing a gap with width $w=(1.5+ 1/2\sqrt{3}) a$ along $\Gamma \text{K}$ direction in a hexagonal crystal. This crystal is created by carving out circular holes with radius $r=0.425a/\sqrt{3}$ through an isotropic Silicon with Lam$\acute{\text{e}}$ constants and mass density $\lambda=44.27G\text{Pa}$, $\mu=80G\text{Pa}$ and $\rho=2329 k\text{g}/\text{m}^{3}$. Slab thickness equals $\text{t} = a/\sqrt{3}$ where $a$ is the lattice constant. Green shaded areas in (a) shows the light cone and the blue regions are the extended modes to  the crystal. The difference in wave number of $S_1$ and $S_2$ equals the wave number of $T_1$, $T_2$ and $T_3$. }
\label{F1}
\end{figure*}

At first we have to address the problem exactly and clearly, then explain how we can solve it. Suppose we have a simultaneous photonic and phononic (phoxonic) waveguide created inside a hexagonal phoxonic crystal slab. This crystal is created by carving out a periodic pattern of circular holes inside a homogeneous isotropic Silicon slab. It has been shown that this crystal may have both photonic and phoninic band gaps in its band structures. Our waveguide as can be seen in the inset of Fig.~\ref{F1}, is created by introducing a path without any holes along $\Gamma$K direction in this crystal. The photonic and phononic band structures of this waveguide are shown in Fig.~\ref{F1}.

Now suppose an optical (photonic) wave is propagating inside this waveguide. We would like to know how this wave evolve in time if an elastic (phononic) wave begins to propagate simultaneously. If we place the crystal slab such that its mid-plane coincides with $xy$ plane of Cartesian coordinate system and the waveguide be along the $x$ axis, then optical and elastic waves may be represented as
\begin{subequations} \label{E1}
\begin{align}
\mathbf{H}(\mathbf{r}, t) &= \tilde{\mathbf{H}} (\mathbf{r}) e^{-\imath (\kappa x - \omega t)},  \label{E1a} \\
\mathbf{u}(\mathbf{r}, t) &= \tilde{\mathbf{u}} (\mathbf{r}) e^{-\imath (q x - \Omega t)},  \label{E1b}
\end{align}
\end{subequations}
where $\mathbf{H}$ is the magnetic field and $\tilde{\mathbf{H}}$ is its amplitude. $\mathbf{u}$ is the displacement and $\tilde{\mathbf{u}}$ is its amplitude. $\kappa$ and $q$ are optical and elastic wave numbers respectively and $\omega$ and $\Omega$ are these waves frequencies. According to Bloch theorem, $\tilde{\mathbf{H}}$ and $\tilde{\mathbf{u}}$ are periodic functions of $x$ with period equals $a$, the lattice constant of the basic crystal.

\section{Master equation}

When optical wave propagates in the waveguide alone and there is no elastic wave, magnetic field satisfies the following equation.
\begin{equation} \label{E2} 
\mathbb{K} \ \mathbf{H}(\mathbf{r}, t) = -\frac{1}{c^2} \frac{\partial ^2}{\partial t^2} \mathbf{H}(\mathbf{r}, t),
\end{equation} 
where $c$ is the light velocity in vacuum , $\mathbb{K} =\nabla \times \big( \eta(\mathbf{r}) \nabla \times (\cdot) \big)$, and $\eta(\mathbf{r}) = 1/ \epsilon_{\text{r}}(\mathbf{r})$ is the relative impermeability of environment. We assume the waveguide to be loss less, that is it is made of materials with real permittivity and impermeability. 
 The magnetic field wave equation is utilized since spatial operator in this equation, $\mathbb{K}$, is hermitian and this property is required in the remaining, while the spatial operator in electric field wave equation is not hermitian. 
 If now an elastic wave begins to propagate in the waveguide, it perturbs the permittivity of the crystal, so the wave equation (Eq. (\ref{E2})) should be modified as
\begin{equation} \label{E3}
\left[ \mathbb{K} + \Delta \mathbb{K}(t) \right] \mathbf{H}'(\mathbf{r}, t) = -\frac{1}{c^2} \frac{\partial ^2}{\partial t^2}\mathbf{H}'(\mathbf{r}, t),
\end{equation} 
where
\begin{align}
\Delta \mathbb{K}(t) &= \Delta \mathbb{K}_{\text{om}}(t) + \Delta \mathbb{K}_{\text{pe}}(t) , \nonumber \\
\Delta \mathbb{K}_{\text{om}}(t) &= \nabla \times \big[ \Delta \eta_{\text{om}} (\mathbf{r}, t) \nabla \times (\cdot) \big] ,\nonumber \\
\Delta \mathbb{K}_{\text{pe}}(t) &= \nabla \times \big[ \Delta \bm{\eta}_{\text{pe}} (\mathbf{r}, t) \nabla \times (\cdot) \big], \nonumber
\end{align} 
 $\Delta \bm{\eta}_{\text{pe}}(\mathbf{r}, t)$ and $\Delta \eta_{\text{om}} (\mathbf{r}, t)$ are the changes in relative impermeability of the waveguide due to photoelasticity and interface displacement respectively. The interface displacement effect is sometimes called opto-mechanical effect in literature. We also use this expression in the remaining. The photoelasticity contribution to the change of relative impermeability equals
\begin{equation} \label{E4}
\Delta \bm{\eta}_{\text{pe}} = \Re \{ \tens{P}: \bm{\varepsilon} \} \Longleftrightarrow \Delta \eta_{ij} = \sum_{k,l = 1}^3 P_{ijkl}  \Re \{\varepsilon_{kl} \},
\end{equation}
in which $ \bm{\varepsilon}= \left[ (\nabla \mathbf{u})^T + \nabla \mathbf{u} \right]/2 $, is the strain wave inside the waveguide and $\tens{P}$ is the fourth rank photo-elastic tensor. The opto-mechanical part which is due to displacement and deformation of the crystal holes equals
\begin{equation} \label{E5}
\Delta \eta_{\text{om}} (\mathbf{r},t)= \eta \left(\mathbf{r}-\Re\{\mathbf{u}(\mathbf{r}, t)\}  \right) - \eta \left(\mathbf{r}\right),
\end{equation}

According to Eqs. (\ref{E1b}) and (\ref{E4}), photoelasticity part can be written as  
\begin{align} \label{E6}
\Delta \bm{\eta}_{\text{pe}} (\mathbf{r}, t) &= \Re \left\{ \bm{v} (\mathbf{r}) e^{-\imath (q x - \Omega t)} \right\} \nonumber \\
&= \bm{v} (\mathbf{r}) e^{-\imath (q x - \Omega t)} + \bm{v}^* (\mathbf{r}) e^{\imath (q x - \Omega t)},
\end{align}
where $\bm{v} (\mathbf{r}) $ is a periodic tensor of $x$ with period $a$. The opto-mechanical part is more complicated, it may be written as (see appendix \ref{AppB})
\begin{equation} \label{E6b}
\Delta \eta_{\text{om}} (\mathbf{r}, t) = \sum_{k=-\infty}^{\infty} \chi_k (\mathbf{r}) e^{\imath k (q x - \Omega t)},
\end{equation}
where $\chi_n (\mathbf{r}),~(\forall n \in \mathbb{Z})$ are periodic functions along $x$ axis with period equals $a$.

In appendix \ref{AppA}, we have proved that $\mathbb{K}$ is a hermitian operator, hence its eigen states form a complete set which we can expand the magnetic field, $\mathbf{H}'(\mathbf{r}, t)$, on them as
\begin{equation} \label{E7}
\mathbf{H}'(\mathbf{r}, t) =  \sum_{\kappa,n} \alpha_{n, \kappa} (t) \tilde{\mathbf{H}}_{n, \kappa} (\mathbf{r}) e^{-\imath (\kappa x - \omega_{n, \kappa} t)},
\end{equation}
where the sum is over all wave numbers in the first Brillouin zone and all eigen states corresponding to each value of wave number. Here we have discretized the wave number just for simplicity. It can be replaced by an integral as 
$ \left( \sum_{\kappa} \longleftrightarrow \frac{a}{\pi} \int_{0}^{\pi/a} \ud \kappa \right) $.
$\alpha_{n, \kappa} (t)$ are time-dependent constants to be determined.

Substituting $\mathbf{H}'(\mathbf{r}, t)$ from Eq. (\ref{E7}) into Eq. (\ref{E3}) yields
\begin{align} \label{E8}
& \Delta \mathbb{K}(t)  \sum_{\kappa,n} \alpha_{n, \kappa} (t) \tilde{\mathbf{H}}_{n, \kappa} (\mathbf{r}) e^{-\imath (\kappa x - \omega_{n, \kappa} t)}    \nonumber \\
& = -\frac{1}{c^2} \sum_{\kappa,n}  \left[ 2 \imath \dot{\alpha}_{n, \kappa} (t) \omega_{n, \kappa}  + \ddot{\alpha}_{n, \kappa} (t) \right]   \tilde{\mathbf{H}}_{n, \kappa} (\mathbf{r}) \nonumber \\
& \qquad \qquad \qquad \qquad \qquad \qquad \qquad \cdot e^{-\imath (\kappa x - \omega_{n, \kappa} t)} \nonumber \\
& \simeq -\frac{1}{c^2}  \sum_{\kappa,n}  2 \imath \dot{\alpha}_{n, \kappa} (t)  \omega_{n, \kappa}  \tilde{\mathbf{H}}_{n, \kappa} (\mathbf{r})  e^{-\imath (\kappa x - \omega_{n, \kappa} t)} .
 \end{align}
In this equation we have assumed variations of $\alpha_{n, \kappa} (t)$ and $\dot{\alpha}_{n, \kappa} (t)$ to be very slow such that $ \ddot{\alpha}_{n, \kappa} (t) \ll \dot{\alpha}_{n, \kappa} (t) \omega_{n, \kappa}$. This assumption is valid in weak coupling regime because in this regime the optical wave state does not change rapidly, that is we can assume $\alpha_{n, \kappa} (t)$ coefficients to be approximately constant in optical wave period time scale, i.e. $ \dot{\alpha}_{n, \kappa} (t) \ll \alpha_{n, \kappa} (t) \omega_{n, \kappa}$.

Now we dot product both sides of Eq. (\ref{E8}) by a specific eigen state of $\mathbb{K}$ such as 
$ \tilde{\mathbf{H}}_{m, \ell} (\mathbf{r})  e^{-\imath (\ell x - \omega_{m, \ell} t)} $ to obtain
\begin{align} \label{E9}
& \frac{1}{d^2 l_x} \int\limits_{V} \sum_{\kappa,n} \alpha_{n, \kappa} (t) \tilde{\mathbf{H}}^*_{m, \ell} (\mathbf{r})  e^{\imath (\ell x - \omega_{m, \ell} t)} \nonumber \\
& \qquad \qquad  \cdot \Delta \mathbb{K}(t) \left[ \tilde{\mathbf{H}}_{n, \kappa} (\mathbf{r}) e^{-\imath (\kappa x - \omega_{n, \kappa} t)} \right] \ud \mathbf{r} \nonumber \\
& \simeq -\frac{1}{c^2}  2 \imath \dot{\alpha}_{m, \ell} (t) \omega_{m, \ell} ,
\end{align} 
where $V$ should grow up to occupy the whole space, $l_x$ is the dimension of $V$ along the waveguide ($x$ axis) and $d$ is the waveguide width. The eigen states are supposed to be normalized, i.e.
\begin{equation*} 
\frac{1}{d^2 l_x} \int\limits_{V} \tilde{\mathbf{H}}^*_{m, \ell} (\mathbf{r})  e^{\imath \ell x} \tilde{\mathbf{H}}_{n, \kappa} (\mathbf{r}) e^{-\imath \kappa x } \ud \mathbf{r} = \delta_{n,m} \delta_{\kappa, \ell} ,
\end{equation*} 
where $\delta_{n,m}$ and $ \delta_{\kappa, \ell}$ are Kronecker delta. 
After Expansion of the left side of Eq. (\ref{E9}), it becomes
\begin{align} \label{E10}
&  \sum_{\kappa,n} \alpha_{n, \kappa} (t) \Big\{ \xi_{\text{pe}_-}^{m,n,\ell, \kappa}   e^{ \imath (\omega_{n, \kappa} -\omega_{m, \ell} + \Omega) t} \nonumber \\
& \qquad \qquad \ +\xi_{\text{pe}_+}^{m,n,\ell, \kappa}   e^{ \imath (\omega_{n, \kappa} -\omega_{m, \ell} - \Omega) t}  \nonumber \\
& \qquad \qquad \ + \sum_{k=-\infty}^{\infty} \xi_{\text{om}_k}^{m,n,\ell, \kappa}  e^{ \imath (\omega_{n, \kappa} -\omega_{m, \ell} - k\Omega) t}  \Big\} \nonumber \\
& \simeq -\frac{ 2 \imath}{c^2}  \dot{\alpha}_{m, \ell} (t) \omega_{m, \ell}
\end{align}
where we have defined the coupling strength between $(\kappa , n)$ and $(\ell , m)$ modes due to photoelastic and optomechanical effects as 
\begin{align} 
&\xi_{\text{pe}_\pm}^{m,n,\ell, \kappa} = \frac{1}{d^2 l_x} \int\limits_{V} \tilde{\mathbf{H}}^*_{m, \ell} (\mathbf{r}) e^{ \imath \ell x} \cdot \nonumber  \\
& \qquad \qquad \quad  \nabla \times \Big[\bar{\bm{v}}(\mathbf{r}) e^{ \pm \imath q x} \nabla \times \left( \tilde{\mathbf{H}}_{n,\kappa} (\mathbf{r}) e^{ -\imath \kappa x} \right) \Big]   \ud \mathbf{r}, \nonumber \\ 
& \xi_{\text{om}_k}^{m,n,\ell, \kappa} = \frac{1}{d^2 l_x} \int\limits_{V} \tilde{\mathbf{H}}^*_{m, \ell} (\mathbf{r}) e^{ \imath \ell x} \cdot \nonumber  \\
& \qquad \qquad \quad  \nabla \times \Big[\chi_k(\mathbf{r}) e^{ \imath k q x} \nabla \times \left( \tilde{\mathbf{H}}_{n,\kappa} (\mathbf{r}) e^{-\imath \kappa x} \right) \Big]   \ud \mathbf{r},\nonumber
\end{align}
in which $\bar{\bm{v}}(\mathbf{r})$ equals $\bm{v}(\mathbf{r}) $ and $\bm{v}^*(\mathbf{r}) $ for $-$ and $+$ subscripts respectively. It is simple to show 
\begin{equation*} 
\xi_{\text{pe}_-}^{m,n,\ell, \kappa} =  \big( \xi_{\text{pe}_+}^{n,m,\kappa,\ell} \big)^* , \quad \xi_{\text{om}_k}^{m,n,\ell, \kappa} = \big( \xi_{\text{om}_{-k}}^{n,m,\kappa,\ell} \big)^*. \nonumber
\end{equation*}
Defining $\beta_{m, \ell} (t) =  \alpha_{m, \ell} (t)  e^{ \imath \omega_{m, \ell}  t}$ and performing some simplification, Eq. (\ref{E10}) becomes
\begin{align} \label{E12}
\dot{\beta}_{m, \ell} (t) &=  \imath \frac{c^2}{2 \omega_{m, \ell} }  \sum_{\kappa,n} \beta_{n, \kappa} (t)  \bigg\{ \xi_{\text{pe}_-}^{m,n,\ell, \kappa}   e^{ \imath \Omega t} \nonumber \\ 
& +\xi_{\text{pe}_+}^{m,n,\ell, \kappa}   e^{ -\imath \Omega t} 
  + \sum_{k=-\infty}^{\infty} \xi_{\text{om}_k}^{m,n,\ell, \kappa}  e^{- \imath  k\Omega t} \bigg\} \nonumber \\ 
&  + \imath \omega_{m, \ell} \beta_{m, \ell} (t) ,
\end{align} 
which is the master equation for the electromagnetic wave evolution. 

\section{Coupling analysis}

Before dealing with the master equation, it is better to investigate the coupling expressions more accurately. 
\subsection{Effect of mode symmetry on coupling}
In general, each of optical and elastic waves inside a phoxonic crystal slab divide into two modes. Optical waves are transverse magnetic like (TM-like) or transverse electric like (TE-like), while elastic waves are horizontal displacement like (HD-like) or vertical displacement like (VD-like). Magnetic field is parallel and vertical to the surface of the crystal slab in the mid-plane of it for TM-like and TE-like modes respectively. On the other hand, displacement field in the mid-plane of the crystal slab is parallel and perpendicular to slab surface for HD-like and VD-like modes respectively. This means if we assume the mid-plane of the slab coincides with $xy$ plane, then in TE-like mode, $H_z(\mathbf{r})$ is an even function and $\mathbf{H}_{xy}(\mathbf{r})$ is an odd function of $z$ , while in TM-like mode, $H_z(\mathbf{r})$ is an odd function and $\mathbf{H}_{xy}(\mathbf{r})$  is an even function of $z$. 
Similarly in VD-like mode, $u_z(\mathbf{r})$ is an even function and $\mathbf{u}_{xy}(\mathbf{r})$ is an odd function of $z$, while in HD-like mode, $u_z(\mathbf{r})$ is an odd function and $\mathbf{u}_{xy}(\mathbf{r})$ is an  even function of $z$. This is summarized in table 1. In this table E stands for even and O stands for odd function.
It should be noted that in positions other than mid-plane of the slab, electromagnetic and displacement fields may have component along any directions.

\begin{table}[!ht]
\renewcommand{\arraystretch}{1.3}
\caption{Fields symmetries with respect to $z$ axis }
\label{T1}
\centering
\begin{tabular}{c c c}
\noalign{\hrule height 1pt}
Mode & Field & Symmetry \\
\hline
TE-like & $\mathbf{H}(\mathbf{r})$  & $\left( \text{O} , \text{O} ,  \text{E} \right)^T$  \\
\hline
TM-like & $\mathbf{H}(\mathbf{r})$  & $\left( \text{E} , \text{E} ,  \text{O} \right)^T$  \\
\hline
VD-like & $\mathbf{u}(\mathbf{r})$  & $\left( \text{O} , \text{O} ,  \text{E} \right)^T$  \\
\hline
HD-like & $\mathbf{u}(\mathbf{r})$  & $\left( \text{E} , \text{E} ,  \text{O} \right)^T$ \\
\noalign{\hrule height 1pt}
\end{tabular}
\end{table}

For the moment we would like to know if the coupling strength between any optical and elastic modes vanishes. For this purpose we have to perform a simple symmetry analysis. We analyze photoelastic and optomechanical coupling between VD-like and TE-like modes and present the results of other modes analysis. 

It is simple to show $\nabla \mathbf{u} $ and $\left( \nabla \mathbf{u} \right)^T$ have
\begin{equation*} 
\begin{pmatrix} 
\text{O} & \text{O} & \text{E}   \\
\text{O} & \text{O} & \text{E}  \\
\text{E} & \text{E} & \text{O}   \\
\end{pmatrix},
\end{equation*} 
symmetry in VD-like mode. So $ \bm{\varepsilon}$ has the same symmetry. Since Voigt notation of Silicon photo-elastic tensor has the following format \cite{Biegelsen},
\begin{equation*} 
\tens{P} =
\begin{pmatrix} 
P_{11} & P_{12} & P_{12} & 0 & 0 & 0   \\
P_{12} & P_{11} & P_{12} & 0 & 0 & 0  \\
P_{12} & P_{12} & P_{11} & 0 & 0 & 0 \\
0 & 0 & 0 & P_{44} & 0 & 0   \\
0 & 0 & 0 & 0 & P_{44} & 0   \\
0 & 0 & 0 & 0 & 0 & P_{44}  
\end{pmatrix} ,
\end{equation*}  
we conclude $\Delta \bm{\eta}_{\text{pe}} $ has the same symmetry as $\nabla \mathbf{u} $. Symmetry of $\nabla \times \mathbf{H}$  is opposite of $\mathbf{H}$, Hence we have,
\begin{align} 
\Delta \bm{\eta}_{\text{pe}} (\nabla \times \mathbf{H}) &\leadsto \left( \text{O} , \text{O} ,  \text{E} \right)^T ,\nonumber \\
\nabla \times \big[ \Delta \bm{\eta}_{\text{pe}} (\nabla \times \mathbf{H})\big] &\leadsto \left( \text{E} , \text{E} ,  \text{O} \right)^T, \nonumber \\
\mathbf{H} \cdot \nabla \times \big[ \Delta \bm{\eta}_{\text{pe}} (\nabla \times \mathbf{H})\big] &\leadsto  \text{O}. \nonumber
\end{align} 
This shows the integrand of $\xi_{\text{pe}_\pm}^{m,n,\ell, \kappa}$ is an odd function of $z$. Since integral is symmetric around $z=0$,  $\xi_{\text{pe}_\pm}^{m,n,\ell, \kappa}$ vanishes for a transition from a TE-like to a TE-like mode through a VD-like mode.

For opto-mechanical coupling, we begin with
\begin{align} 
\mathbf{r} &\leadsto \left( \text{E} , \text{E} ,  \text{O} \right)^T ,\nonumber \\
\mathbf{u}(\mathbf{r}) &\leadsto \left( \text{O} , \text{O} ,  \text{E} \right)^T, \nonumber \\
\mathbf{r} - \mathbf{u}(\mathbf{r}) &\leadsto \left( \text{N} , \text{N} ,  \text{N} \right)^T , \nonumber
\end{align}
where N stands for not even nor odd. We know $\eta (\mathbf{r})$ is an even function of $z$ but with respect to Eq. (\ref{E5}) we conclude $\Delta \eta_{\text{om}} $ is not an even nor odd function of $z$. Hence $\xi_{\text{om}_k}^{m,n,\ell, \kappa}$ has value for VD-like elastic wave coupled with both TE-like and TM-like modes of electromagnetic wave. The coupling analysis results between different modes are collected in table 2.

\begin{table}[!ht]
\renewcommand{\arraystretch}{1.3}
\caption{Coupling analysis results }
\label{T2}
\centering
\begin{tabular}{c c c c c}
\noalign{\hrule height 1pt}
$\text{Mode-i}^{\dagger}$  & $\text{Mode-f}^{\ddagger}$ & $\text{MechMode}^{\star}$ & $\xi_{\text{pe}_{\pm}}^{m,n,\ell, \kappa}$ & $\xi_{\text{om}_k}^{m,n,\ell, \kappa}$ \\
\hline
TE-like & TE-like  & VD-like & 0  & $\text{NZ}^{\diamond}$\\
\hline
TE-like & TM-like  & VD-like & NZ  & NZ\\
\hline
TM-like & TE-like  & VD-like & NZ  & NZ\\
\hline
TM-like & TM-like  & VD-like & 0  & NZ\\
\hline
TE-like & TE-like  & HD-like & NZ  & NZ\\
\hline
TE-like & TM-like  & HD-like & 0  & 0\\
\hline
TM-like & TE-like  & HD-like & 0  & 0\\
\hline
TM-like & TM-like  & HD-like & NZ  & NZ\\
\noalign{\hrule height 1pt}
\end{tabular}
\raggedright{\footnotesize{$\dagger~$Initial electromagnetic state.} \
\footnotesize{$\ddagger~$Final electromagnetic state.}\ 
\footnotesize{$\star~$Elastic wave mode.} \quad \footnotesize{$\diamond~$Not necessarily zero.} \\}
\end{table}

\subsection{Effect of wave number on coupling}
Another fact about coupling strengths we can conclude by carefully investigating their expressions is that they become zero if initial optical wave number plus elastic wave number does not equal the final optical wave number, i.e. $\xi_{\text{pe}_{\pm}}^{m,n,\ell, \kappa} = 0$ for $\kappa \mp q \ne \ell$ and $\xi_{\text{om}_{k}}^{m,n,\ell, \kappa} = 0$ for $\kappa + kq \ne \ell $.
To prove this claim, we consider that $\bar{\bm{v}} (\mathbf{r})$, $ \chi_k (\mathbf{r})$ and $\tilde{\mathbf{H}}$ are periodic functions of $x$ with period equals $a$. So product of functions other than exponential terms in coupling integrand are perodic which can be expanded as a Fourier series,
\begin{equation*} 
\sum_{h=-\infty}^{\infty} f_h(y,z) \exp \left(\imath h \frac{2\pi}{a} x \right).
\end{equation*} 
Inserting this series into coupling expressions we reach
\begin{align} 
\xi_{\text{pe}_\pm}^{m,n,\ell, \kappa} &= \sum_{h=-\infty}^{\infty} \mathcal{I}_h=  \sum_{h=-\infty}^{\infty} \frac{1}{d^2 l_x} \nonumber \\
&  \int\limits_{V} f_h(y,z) e^{ \imath( \ell \pm q - \kappa) x}  \exp \left(\imath h \frac{2\pi}{a} x \right)   \ud \mathbf{r} \nonumber \\
&= \mathcal{I}_{h_0}, \nonumber
\end{align}  
where  $2\pi h_0/a + (\ell \pm q - \kappa)=0$. 
The same can be written for $\xi_{\text{om}_k}^{m,n,\ell, \kappa}$.

\subsection{Coupling with leaky modes }

Leaky modes inside a photonic crystal slab waveguide can be divided into two categories. 1) Modes which do not confine inside the slab and extend along normal axis of the slab ($z$ axis). These modes lie above the light cone in the waveguide band structure. 2) Modes which are confined in the crystal slab but propagate along $y$ axis and do not stay in the waveguide. Power of both of these modes are distributed in the whole space, i.e. their power density is zero because, 
\begin{equation*} 
\frac{1}{d^2 l_x} \int\limits_{V} |\tilde{\mathbf{H}}_{n, \kappa} (\mathbf{r}) |^2 \ud \mathbf{r} = 1 , \quad (V= \text{ whole space}).
\end{equation*} 
Since guided elastic waves are confined in the waveguide, integral of coupling expressions are taken over a finite region around the waveguide, hence this integral vanishes.

\section{Optical wave evolution}
Now we are at a point to solve the master equation. 
But before that, it is better to find a physical interpretation for it which gives us an insight of what is happening during the interaction. This enables us to further simplify the equation. In general, Eq. (\ref{E10}) shows that the rate of change of a single optical mode amplitude, i.e. $\alpha_{m, \ell} $, is directly related to how strong other modes couple to elastic wave during transition to that mode, i.e. how large the coupling coefficients related to transition from other modes to that specific mode are. By investigating this equation, we find two mechanisms through which other modes can transit to a specific mode, $(\ell , m)$, via photoelastic effect. Absorbing a phonon with frequency $\Omega$ and wave vector $q$ which satisfies $\kappa + q =\ell$, or emitting a phonon with frequency $\Omega$ and wave vector $q$ which satisfies $\kappa - q =\ell$. The interface displacement effect introduces infinite mechanisms which can cause transition to the mode $(\ell , m)$. In these mechanisms, the optical wave absorbs or emits $k$ (any positive integer) phonons with frequency $\Omega$ and wave vector $q$, provided that $\kappa \pm kq =\ell$, where $+$ and $-$ relate to absorption and emission respectively.

 Another important fact that can be concluded from Eq. (\ref{E10}) is that in weak coupling regime, the greater the exponential power, the fewer the related transition contribute to the amplitude of the intended mode, $(\ell , m)$. In other words, the transitions in which the frequency difference between the initial and final modes approximately equals the elastic wave frequency times $k$ (number of phonons absorbed or emitted) contribute dominantly in the amplitude of the final mode. This means we can just keep those transitions with exponential power $\lesssim \Omega$ to obtain a simplified form of the master equation.

As a simple example, suppose the electromagnetic wave is initially in a pure (single frequency) state such as $(\kappa_0, n_0)$.
As stated above, we ignore transition to modes with large frequency differences ($\gg \Omega$), so the modes with nearest frequency to the initial mode are kept in the master equation. 
Since no interaction occurs with leaky modes, a few number of states contribute in the equation.

The simplest form of Eq. (\ref{E12}) would be obtained if only two pure states ($S_1$ and $S_2$) existed in the equation. For $\kappa_1 + q = \kappa_2$, it becomes
\begin{subequations} \label{E13}
\begin{align} 
\dot{\beta}_{1} (t) &=  \imath \omega_{1} \beta_{1} (t) + \imath \zeta_{1,2}  e^{ -\imath \Omega t} \beta_{2} (t)  , \label{E14a} \\
\dot{\beta}_{2} (t) &=  \imath \zeta_{2,1} e^{ \imath \Omega t} \beta_{1} (t) + \imath \omega_{2} \beta_{2} (t)   , \label{E14b}\\
\beta_1 (0) &= 1, \quad \beta_2 (0) =0, \nonumber
\end{align} 
\end{subequations}
where $\zeta_{1,2} = c^2 \big(\xi_{\text{pe}_+}^{1,2} + \xi_{\text{om}_1}^{1,2} \big) / 2 \omega_{1}$ and $\zeta_{2,1} = c^2 \big( \xi_{\text{pe}_-}^{2,1} + \xi_{\text{om}_{-1}}^{1,2} \big)/ 2 \omega_{2}$.
Solving this system of equations with Laplace transform technique, we obtain 
\begin{align}
\beta_1 (t) &= \frac{\exp \big[ -\imath (\Omega -\omega_1-\omega_2)t /2 \big] }{\sqrt{4 \zeta_{1,2} \zeta_{2,1}+(\Omega +\omega_1- \omega_2)^2}} \bigg\{ \imath (\Omega +\omega_1- \omega_2)  \nonumber \\
& \ \cdot \sin \left[ \left( \sqrt{4  \zeta_{1,2} \zeta_{2,1}+(\Omega +\omega_1- \omega_2)^2}\right) t/2 \right] \nonumber \\
& \  +\sqrt{4 \zeta_{1,2} \zeta_{2,1}+(\Omega +\omega_1- \omega_2)^2}  \nonumber \\
& \ \cdot \cos \left[ \left( \sqrt{4  \zeta_{1,2} \zeta_{2,1}+(\Omega +\omega_1- \omega_2)^2}\right) t/2 \right] \bigg\}, \displaybreak \nonumber \\
\beta_2 (t) &= \frac{2 \imath  \zeta_{2,1} \exp \big[ \imath (\Omega +\omega_1+\omega_2)t /2 \big] }{\sqrt{4 \zeta_{1,2} \zeta_{2,1}+(\Omega +\omega_1- \omega_2)^2}} \nonumber \\
& \  \cdot \sin \left[ \left( \sqrt{4  \zeta_{1,2} \zeta_{2,1}+(\Omega +\omega_1- \omega_2)^2}\right) t/2 \right], \nonumber
\end{align} 
which yields
\begin{align} \label{E14}
\alpha_1 (t) &= \exp \big[ -\imath \Delta t /2 \big]   \left( \imath \frac{\Delta}{\varpi} \sin \left[ \varpi t/2 \right] + \cos \left[ \varpi t/2 \right] \right) , \nonumber \\
\alpha_2 (t) &= \exp \big[ \imath \Delta t /2 \big] \imath \frac{2 \zeta_{2,1}  }{\varpi} \sin \left[ \varpi t/2 \right], 
\end{align} 
where $\Delta = \Omega +\omega_1- \omega_2$ and $\varpi = \sqrt{4  \zeta_{1,2} \zeta_{2,1}+\Delta^2}$.

We know $\omega_1$ and $\omega_2$ are at least $4-5$ orders of magnitude greater than $\Omega$, hence if $S_2$ is the only mode which its frequency, $\omega_2$, approximately equals $\omega_1$ such that $\Delta \sim \Omega$ and the coupling is weak, i.e. $|\zeta_{1,2}| \ll \omega_1$, then omitting other modes from the master equation, as we did, does not enter large errors into our calculations.
This is because for other modes (for example mode $S_j$), $\Delta / \varpi \rightarrow 1$ and $\zeta_{j,1} / \varpi \rightarrow 0$, that is $\alpha_j (t) \simeq 0$.

\begin{figure*}[!ht]
\centering
{\includegraphics[width=3.4in]{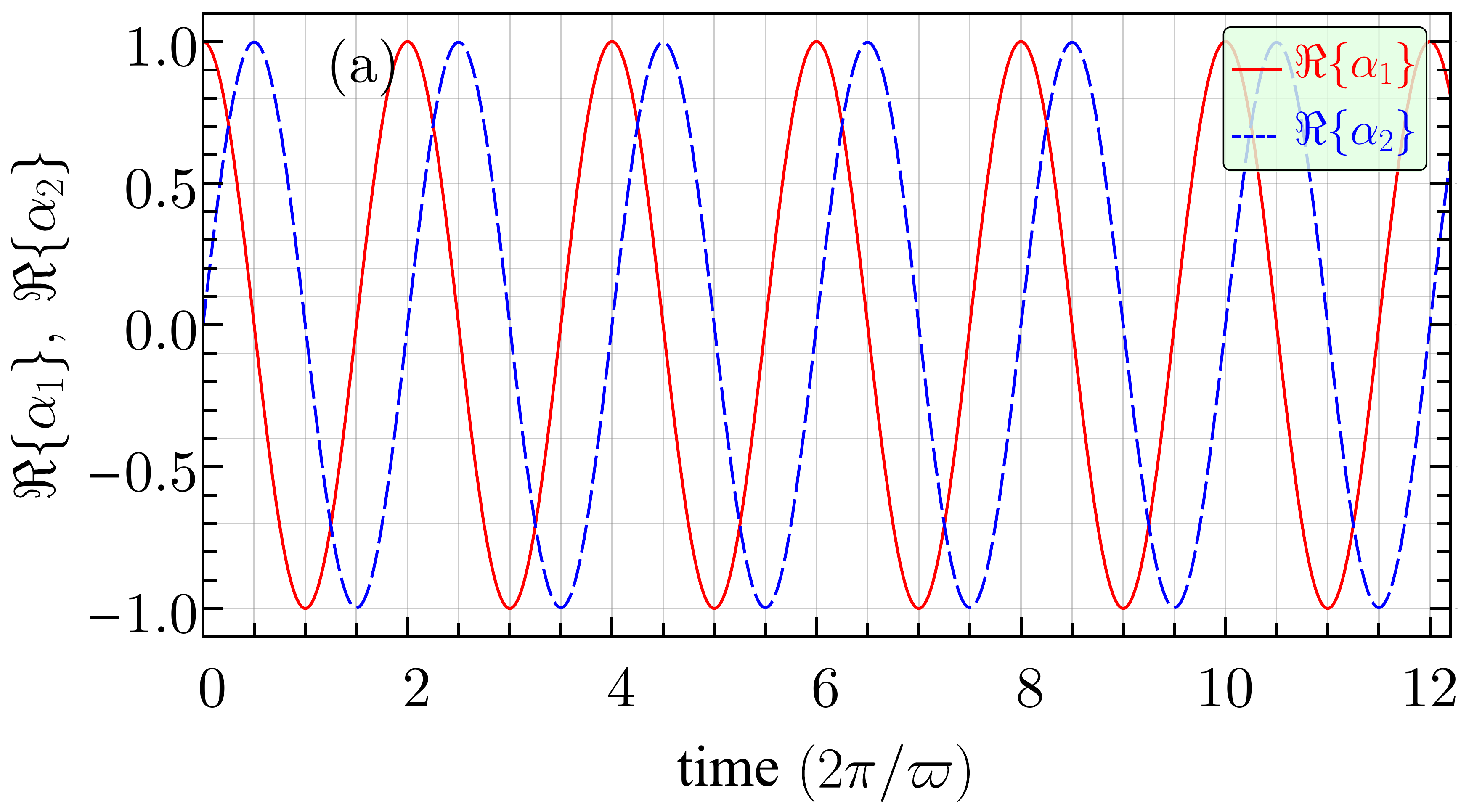}\label{F2a}}
\hfill
{\includegraphics[width=3.4in]{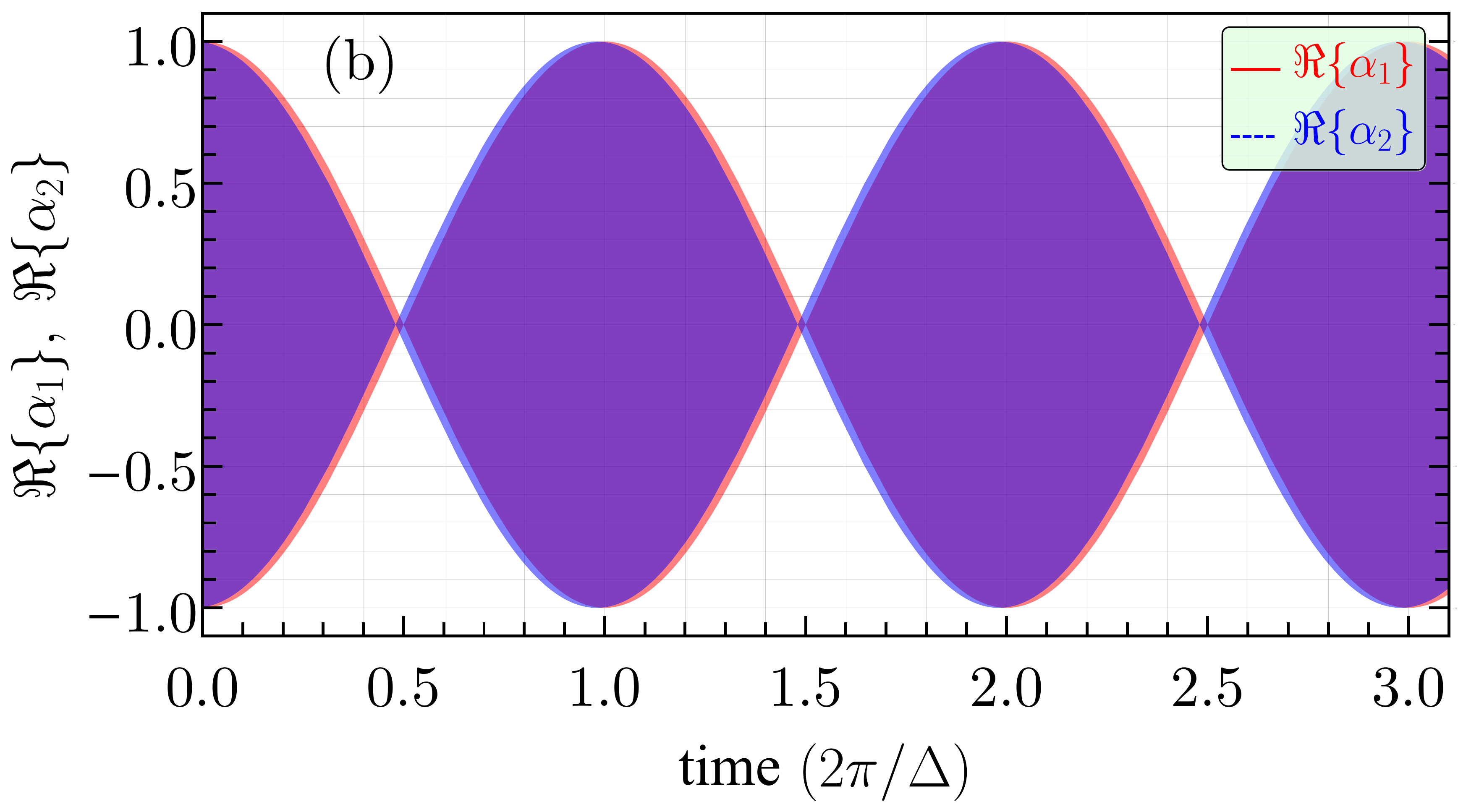}\label{F2b}} \\
{\includegraphics[width=3.4in]{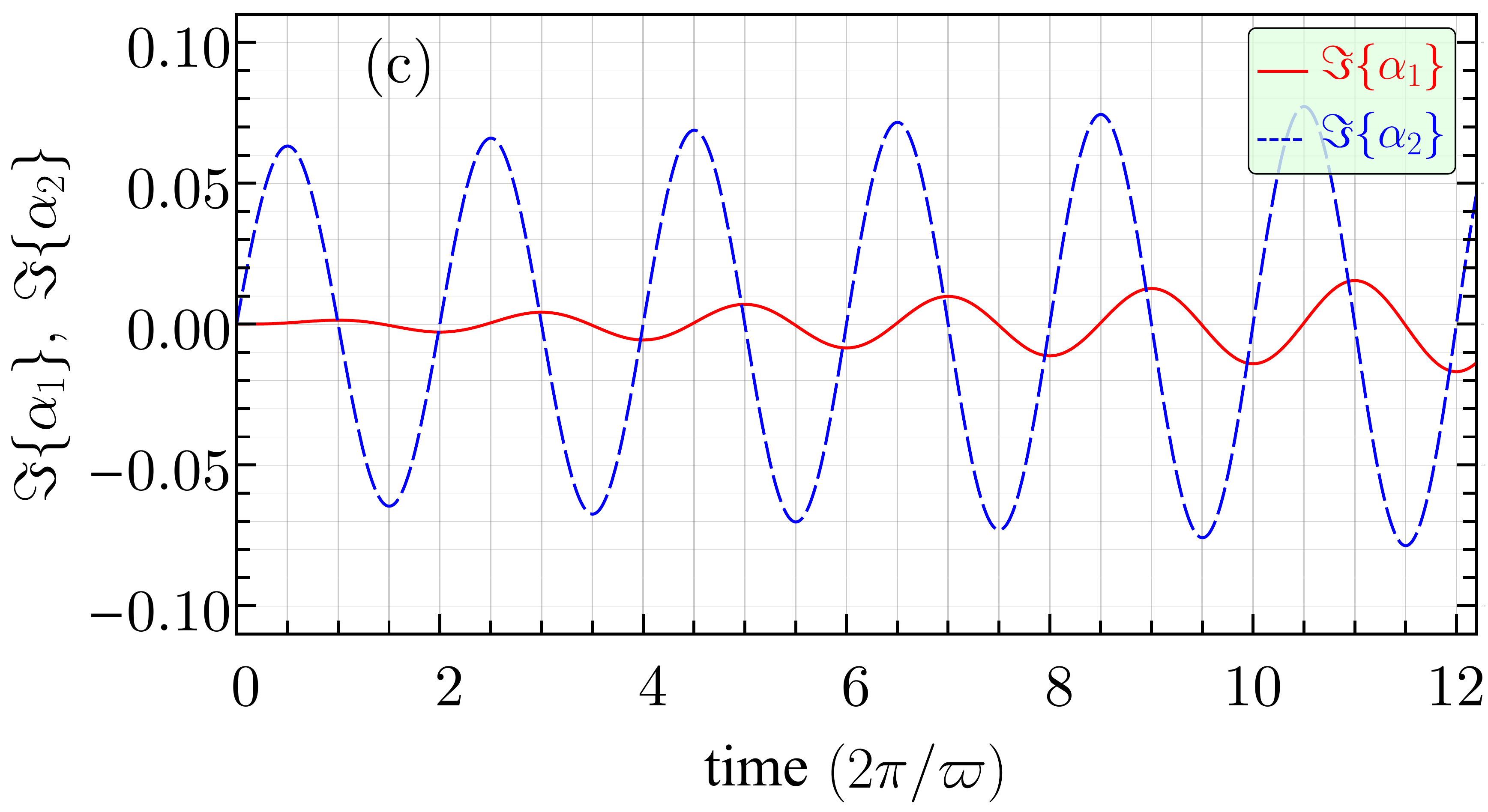}\label{F2c}}
\hfill
{\includegraphics[width=3.4in]{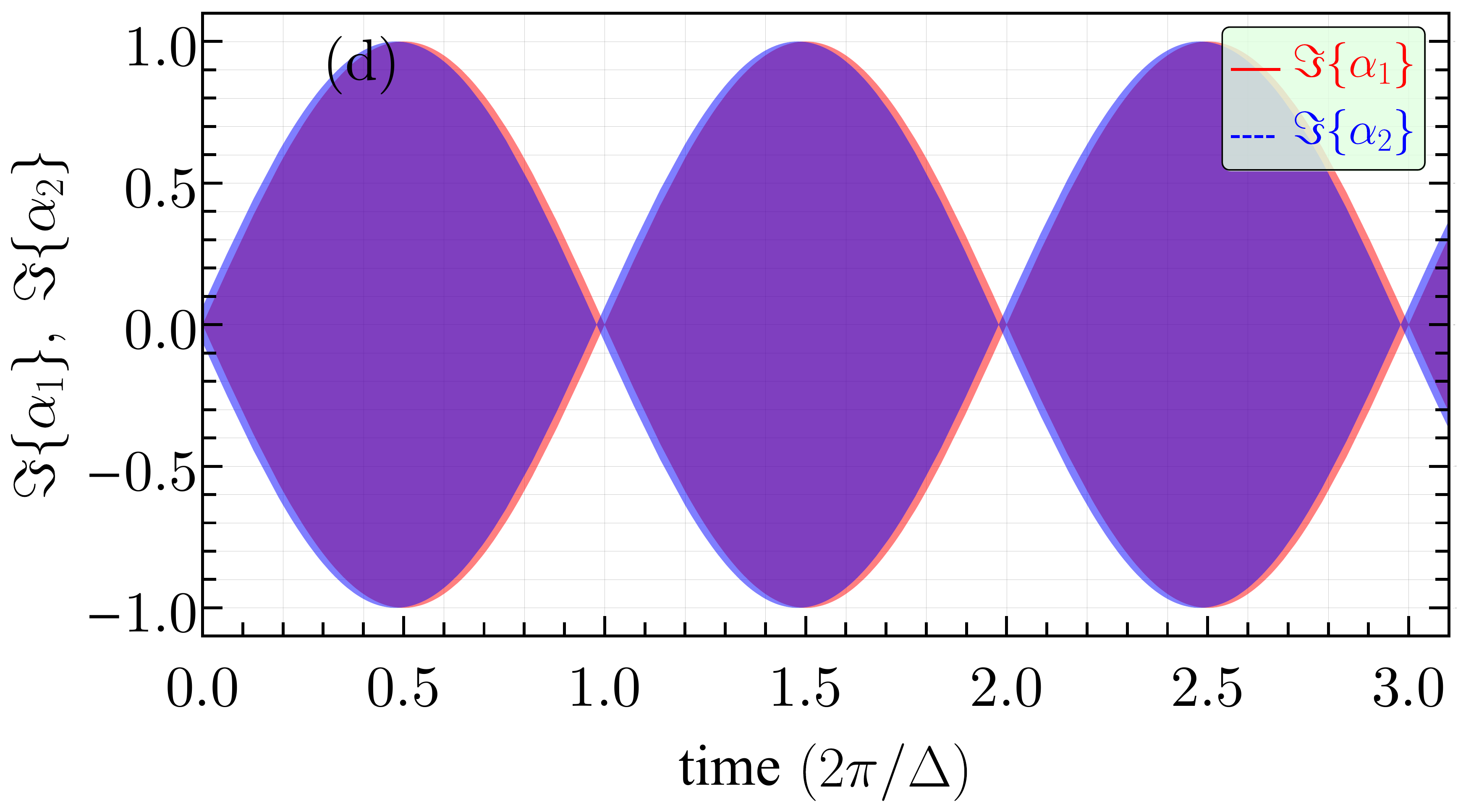}\label{F2d}}
\caption{Variations of $\alpha_1(t)$ and $\alpha_2(t)$ in short ((a) and (c)) and long ((b) and (d)) time period. These coefficients have both fast and slow oscillations. As shown in (a) and (c) the frequency of fast oscillations is $\varpi/2$ while the frequency of slow oscillations equals $\Delta/2$ as can be seen in (b) and (d). Since $\varpi \simeq 9.04 \times 10^6 \Delta$, fast oscillations in (b) and (d) are not distinguishable.  }
\label{F2}
\end{figure*}

As a practical example consider the phoxonic waveguide introduced in section \ref{S2} (Fig \ref{F1}). Suppose the optical wave is initially in state $S_1$  and at $t=0$ it begins to interact with the single frequency elastic wave, $T_2$. Since there is a mode ($S_2$) with frequency $\omega_2=\omega_1=0.286  \times 2\pi c / a \ \text{rad}/\text{s}$, transition to this mode dominates and other modes of the band structure can be ignored. Coupling constants equal $\xi_{\text{pe}_+}^{1,2} \simeq (2.55 - \imath 8)\times 10^{-3}/a^2$ and $\xi_{\text{om}_1}^{1,2} \simeq (3.4+ \imath 103)\times 10^{-3}/a^2$ for,
\begin{equation*} 
\sqrt{\frac{1}{d^2 l_x} \int\limits_{V} |\mathbf{u}|^2 \ud \mathbf{r}} = 0.1 a .
\end{equation*} 
Inspecting Fig. \ref{F1} we find the frequency of mode $T_2$ to be $\Omega_2 = 3654 \times 2\pi /a \ \text{rad}/\text{s}$. This means $\omega_1 \simeq 23481 \Omega_2$ and $\varpi \simeq 9.04 \times 10^6 \Delta$. Using these values we have plotted variations of $\alpha_1 (t)$ and $\alpha_2(t)$ in Fig. \ref{F2}. Parts (a) and (c) of this figure show their variations in short time period while parts (b) and (d) illustrate their long-term behavior.
According to this figure, electromagnetic wave periodically oscillates between states $S_1$ and $S_2$. This oscillation may be interpreted as absorption and emission of phonons from and to elastic wave
by photons of the optical wave periodically.

\section{Conclusion}
A master equation was developed for studying optical wave evolution due to interaction with elastic wave in a phoxonic crystal waveguide. By symmetry analysis, we showed some types of electromagnetic and elastic waves may not interact. We also showed it is possible to change the mode of electromagnetic wave with VD-like elastic waves. By solving the master equation, we illustrated a single frequency electromagnetic wave may experience a periodic transition between its initial state and another state which its frequency is very near to the initial frequency. The rate of this transition depends on the coupling strength between the electromagnetic and elastic waves. The stronger the coupling, the faster this transition occurs.

\begin{acknowledgement}
This work was funded by the Iranian National Science Foundation (INSF) under grant no. 93026841.
\end{acknowledgement}

\appendix

\section{Fourier series expansion of $\Delta \eta_{\text{om}}(\mathbf{r},t)$} \label{AppB}
At first we introduce a new independent variable $\lambda = q x -\Omega t$. Using this variable and Eq. (\ref{E1b}), $\mathbf{u}$ could be written as
\begin{equation*} 
\mathbf{u} (\mathbf{r}, \lambda) = \tilde{\mathbf{u}} (\mathbf{r}) e^{-\imath \lambda},
\end{equation*} 
Since $\tilde{\mathbf{u}} (\mathbf{r})$ and $\eta (\mathbf{r})$ are periodic functions of $x$ with period $a$ and $e^{-\imath \lambda}$ is a periodic function of $\lambda$ with period $2\pi$, according to Eq. (\ref{E5}), $\Delta \eta_{\text{om}} (\mathbf{r},\lambda)$ is also a periodic function of $x$ and $\lambda$. Hence it can be written as 
\begin{equation*}
\Delta \eta_{\text{om}} (\mathbf{r}, \lambda) = \sum_{k=-\infty}^{\infty} \chi_k (\mathbf{r}) e^{\imath k \lambda},
\end{equation*} 
where $\chi_k (\mathbf{r})$ are periodic functions of $x$. Now if we replace $\lambda$ by its value, $\Delta \eta_{\text{om}} (\mathbf{r}, t)$ becomes 
\begin{equation*}
\Delta \eta_{\text{om}} (\mathbf{r}, t) = \sum_{k=-\infty}^{\infty} \chi_k (\mathbf{r}) e^{\imath k ( q x -\Omega t)}.
\end{equation*}

\section{Proof of hermiticity of $\mathbb{K}$} \label{AppA}
 To prove $\mathbb{K} =\nabla \times \big( \eta(\mathbf{r}) \nabla \times (\cdot) \big)$, with real function $\eta(\mathbf{r})$, is a hermitian operator, we have to show
\begin{equation} \label{AE1}
\left[ \mathbf{G}(\mathbf{r}) , \mathbb{K} \mathbf{F} (\mathbf{r}) \right] = \left[\mathbb{K} \mathbf{G}(\mathbf{r}) , \mathbf{F} (\mathbf{r}) \right],
\end{equation} 
for arbitrary vector fields $\mathbf{F}(\mathbf{r})$ and $\mathbf{G}(\mathbf{r})$, where
\begin{equation*} 
 [\mathbf{G}(\mathbf{r}) , \mathbf{F} (\mathbf{r}) ] = \frac{1}{d^2 l_x} \int\limits_{V} \mathbf{G}^*(\mathbf{r}) \cdot \mathbf{F} (\mathbf{r}) \ud \mathbf{r},
\end{equation*} 
represents the inner product of $\mathbf{G}(\mathbf{r})$ and $\mathbf{F}(\mathbf{r})$. We begin with the left side of Eq. (\ref{AE1}) and write,
\begin{align} 
\left[ \mathbf{G}, \mathbb{K} \mathbf{F}  \right] &=  \frac{1}{d^2 l_x} \int\limits_{V} G_x^* \big[\eta ^y \left(F_y^x-F_x^y\right)+\eta  \left(F_y^{x,y}-F_x^{y,y}\right)\nonumber \\
& -\eta ^z \left(F_x^z-F_z^x\right)-\eta  \left(F_x^{z,z}-F_z^{x,z}\right) \big] \nonumber \\ 
&+G_y^* \big[-\eta ^x \left(F_y^x-F_x^y\right)-\eta  \left(F_y^{x,x}-F_x^{x,y}\right) \nonumber \\
&+\eta ^z \left(F_z^y-F_y^z\right)+\eta  \left(F_z^{y,z}-F_y^{z,z}\right)\big] \nonumber \\
&+G_z^* \big[\eta ^x \left(F_x^z-F_z^x\right)+\eta  \left(F_x^{x,z}-F_z^{x,x}\right) \nonumber \\
&-\eta ^y \left(F_z^y-F_y^z\right)-\eta  \left(F_z^{y,y}-F_y^{y,z}\right)\big] \ud \mathbf{r}.
\end{align} 
In this equation $f^{i,j}= \partial^2 f /\partial i \partial j , \ (i,j = x,y,z) $.
Utilizing integration by parts, we can convert this integral to 
\begin{align} 
 &  \frac{1}{d^2 l_x} \int\limits_{V}F_x \big[ \eta ^y \left(G_y^x-G_x^y\right) -\eta ^z \left(G_x^z-G_z^x\right) \nonumber \\
& -\eta  \left(G_x^{z,z}-G_z^{x,z}\right)+\eta  \left(G_y^{x,y}-G_x^{y,y}\right) \big] ^* \nonumber \\ 
&+F_y \big[\eta ^z \left(G_z^y-G_y^z\right)+\eta  \left(G_z^{y,z}-G_y^{z,z}\right) \nonumber \\
&-\left(G_y^x-G_x^y\right) \eta ^x-\eta  \left(G_y^{x,x}-G_x^{x,y}\right)\big]^* \nonumber \\
&+F_z \big[-\left(G_z^y-G_y^z\right) \eta ^y-\eta  \left(G_z^{y,y}-G_y^{y,z}\right) \nonumber \\
&+\left(G_x^z-G_z^x\right) \eta ^x+\eta  \left(G_x^{x,z}-G_z^{x,x}\right)\big]^* \ud \mathbf{r} \nonumber \\
& = \left[ \mathbb{K} \mathbf{G},  \mathbf{F}  \right].
\end{align} 
For this conversion we have assumed vector fields vanish as $\mathbf{r} \rightarrow \infty$.

% BibTeX users please use
\bibliographystyle{ieeetr}
\bibliography{mybibfile}
%\begin{thebibliography}{}
%Author, Journal \textbf{Volume,} (year) page numbers.
%Author, \textit{Book title} (Publisher, place year) page numbers
% etc
%\end{thebibliography}

\end{document}